\documentclass[twocolumn,prb,showpacs]{revtex4}
\usepackage{graphicx}
\usepackage[dvips]{color}
\usepackage{amssymb}
\newcommand{\GSO}{$\rm Gd_2Sn_2O_7$}
\newcommand{\GTO}{$\rm Gd_2Ti_2O_7$}
\newcommand{\ETO}{$\rm Er_2Ti_2O_7$}

\begin{document}
\title{Magnetic excitations in the XY-pyrochlore antiferromagnet Er$_2$Ti$_2$O$_7$}
\author{S.\,S.\,Sosin}
\affiliation{P.\,L.\,Kapitza Institute for Physical Problems RAS, 119334 Moscow, Russia}
\author{L.\,A.\,Prozorova}
\affiliation{P.\,L.\,Kapitza Institute for Physical Problems RAS, 119334 Moscow, Russia}
\author{M.\,R.\,Lees}
\affiliation{Department of Physics, University of Warwick, Coventry CV4 7AL, UK}
\author{G.\,Balakrishnan}
\affiliation{Department of Physics, University of Warwick, Coventry CV4 7AL, UK}
\author{O.\,A.\,Petrenko}
\affiliation{Department of Physics, University of Warwick, Coventry CV4 7AL, UK}
\date{\today}
\begin{abstract}
The XY-pyrochlore antiferromagnet \ETO\ is studied by heat
capacity measurements and electron spin resonance spectroscopy
performed on single crystal samples. The magnetic phase diagrams
are established for two directions of applied field, $H\parallel
[100]$ and $H\parallel [111]$. In the magnetically ordered phase
observed below $T_N=1.2$~K, the magnetic excitation spectrum
consists of a Goldstone mode acquiring an isotropic gap in an
applied field, and another mode with a gap softening in the
vicinity of a field-induced phase transition. This second-order
transition takes place at a critical field $H_c$ above which the
magnetization process is accompanied by a canting of the magnetic
moments off their local ``easy-planes''. The specific heat curves
for $H\parallel [100]$ ($H\gg H_c$) are well described by a model
presuming a single dispersionless excitation mode with the energy
gap obtained from the spectroscopic measurements.
\end{abstract}

\maketitle
\section{Introduction}
Compounds with strong geometrical frustration of magnetic bonds
have attracted much attention due to the unusual properties of
their manifold ground states and their peculiar spin dynamics. For
some systems the degeneracy of the ground state is infinite (in
the nearest neighbor exchange approximation), which leads to (i)
delayed magnetic ordering and a wide temperature interval with a
short-range correlated state, the so called cooperative
paramagnet,\cite{villain} and to (ii) an enhanced role for the
weaker interactions in the eventual formation of a long-range
order state at low temperature. Different mechanisms for lifting
the macroscopic degeneracy and selecting a specific ground state
is a matter of individual consideration for each system. In the
case of pyrochlore magnets, (nearly) fully ordered states,
cooperative paramagnetic states, spin-glass and spin-ice states
have all been observed in different
materials.\cite{Gardner_review}

Experimental investigations of excitation spectra often provide
the only viable method of determining a specific set of
interactions to be taken into account for an individual magnetic
compound. For example, in a Heisenberg pyrochlore magnet \GSO,
where the four-sublattice structure with ${\bf k}=0$ (a so-called
plane cross or a Palmer-Chalker state) is found at low
temperature,\cite{wills} the spin-wave calculations performed for
this type of ordering with known microscopic parameters of
exchange, dipolar interactions and a single-ion
anisotropy~\cite{gingras2,sosin0} are in perfect agreement with
the experimentally obtained picture, thus demonstrating the
validity of this description.

Unlike the Gd-based Heisenberg pyrochlores, erbium titanate \ETO\
has a strong local XY-type anisotropy of magnetic moments due to
the large orbital momentum of Er$^{3+}$ ions ($L=6$) which is not
quenched by the crystal field. The Curie-Weiss temperature in this
system obtained from the susceptibility data points to an
antiferromagnetic nearest-neighbor exchange interaction. Depending
on the temperature interval of data fitting, the value of
$\Theta_{CW}$ varies from -13~K in the higher temperature limit
80-300~K~\cite{dasgupta} to -22~K if determined from data
collected at temperatures below 50~K.\cite{bramwell} \ETO\ is
known to undergo a second order transition to a long-range ordered
state at about 1.2~K.\cite{blote,ramirez,champion} A spherical
neutron polarimetry experiment indicates that the magnetic
structure is formed from six domains, all of which are equally
occupied.\cite{Poole} In addition, recent high resolution neutron
diffraction measurements detect the coexistence of short and
long-range order.\cite{gaulin} Low-lying excitation modes ungapped
in zero field and near a critical field of about 15~kOe are also
observed in this experiment. Muon spin-relaxation measurements
show non-vanishing spin dynamics as the temperature approaches
zero,\cite{Lago} in contrast to the expectations for a
conventional magnet.

The non-coplanar structure appearing below $T_N$ is presumed to be
induced by fluctuations,\cite{champion,Chempion_Holdsworth} but
this model predicts a strong first-order phase transition, while
experimentally it is found to be continuous. A mean field model
which includes single-ion anisotropy and anisotropic exchange
interactions is capable of producing the required long-range
order,\cite{gingras} however thermal or quantum fluctuations are
expected to be important in this material. Polarized neutron
diffraction measurements~\cite{Cao} suggest that in addition to
crystal field parameters, an anisotropic molecular field tensor
must be included in order to reproduce the experimental values of
the local susceptibility tensor.

We present the results of an extensive electron spin resonance
study of \ETO, which is used as a high-resolution probe of the
$q=0$ energy structure for different directions of an applied
field. A Goldstone mode is observed and its linear increase with
an applied magnetic field is traced. Additionally, we have
detected a gapped mode softening in the vicinity of a critical
field. The origin of both types of oscillation is discussed.
Specific heat measurements were also carried out for $H\parallel
[100]$ and $H\parallel [111]$ to complement the previously
reported results~\cite{gaulin} for $H\parallel [110]$. The
calorimetry results are compared with the ESR and inelastic
neutron scattering data~\cite{gaulin} and an overall agreement
between the different experimental methods is demonstrated.

\section{Experimental procedures}
A single crystal sample of \ETO\ was grown by the floating zone
technique.\cite{krishna} Small thin plates of a characteristic
size $1\times 1\times 0.2$~mm$^3$ (about 1~mg in mass) containing
the (110)-plane were cut out of the original larger sample. The
plates were mounted in such a way that the magnetic field was
applied in the sample plane in order to minimize the
demagnetization effect. Specific heat was measured using a Quantum
Design Physical Property Measurement System calorimeter equipped
with a $^3$He option and a 90~kOe cryomagnet.

ESR measurements were carried out using a home-made transmission
type spectrometer built on a $^3$He-cryostat with a base
temperature of 0.45~K. The lowest eigen-frequency of the resonant
cavity was 25~GHz. The highest frequency was limited to about
70~GHz by the so-called ``size effect'' {\it i.e.} the condition
that the half-wavelength of the microwave radiation inside the
sample is of the same order as the sample size. For the pyrochlore
magnets, this limit is especially rigid because of their large
magnetic permeability. It has been shown in case of another
pyrochlore, \GTO, that above this frequency limit a parasitic
absorption (not directly related to magnetic excitation spectrum
of the samples) can disguise its resonance pattern.\cite{sosin}

\section{Experimental results}
\begin{figure}
\centerline{\includegraphics[width=\columnwidth]{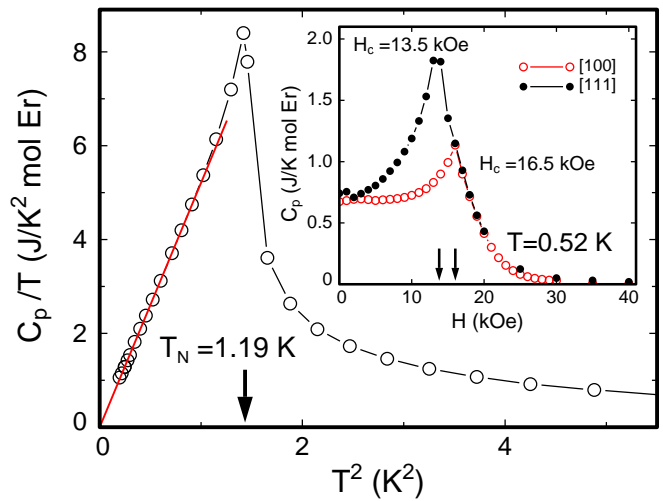}}
\caption{(Color online). Specific heat divided by temperature of
\ETO\ in zero field. Solid line is $C_p/T\propto T^2$ fit for the
low-temperature part of the curve. The inset shows the field
dependencies of the specific heat at $T=0.52$~K: $H\parallel
[100]$ -- {\Large $\circ$}, $H\parallel [111]$ -- {\Large
$\bullet$}.} \label{spheat}
\end{figure}

The temperature dependence of the specific heat in zero external
magnetic field is shown in Fig.~\ref{spheat}. It demonstrates a
sharp anomaly at $T_N=1.2$~K in close agreement with the
previously obtained results. This peak corresponding to a
second-order phase transition into a magnetically ordered state is
followed by $C_p\propto T^3$ drop of specific heat on decreasing
the temperature. The magnetic entropy of the system obtained by
integrating the $C_p/T$ curve with the phonon contribution
subtracted (see below for details) saturates at the value of
$R\ln{2}$. This points to the effective $S=1/2$ pseudo-spin
character of the magnetic Er$^{3+}$ ions originating from the
double degeneracy of its ground state in a crystal field.

In an applied magnetic field the transition shifts to lower
temperatures and becomes difficult to detect in the $C(T)$ curves.
The transition is also traced by recording the field dependence of
the specific heat at constant temperature. The $C(H)$ dependence
measured for two different directions of the external magnetic
field, $H\parallel [100]$ and $[111]$, shows that the peak
anomalies are easily detectable (see the inset to
Fig.~\ref{spheat}) and that their behaviors are quite anisotropic
in nature with the transitions occurring at $H_c^{[100]} =16.5\pm
0.5$~kOe and $H_c^{[111]}=13.5\pm 0.5$~kOe.

\begin{figure}
\centerline{\includegraphics[width=\columnwidth]{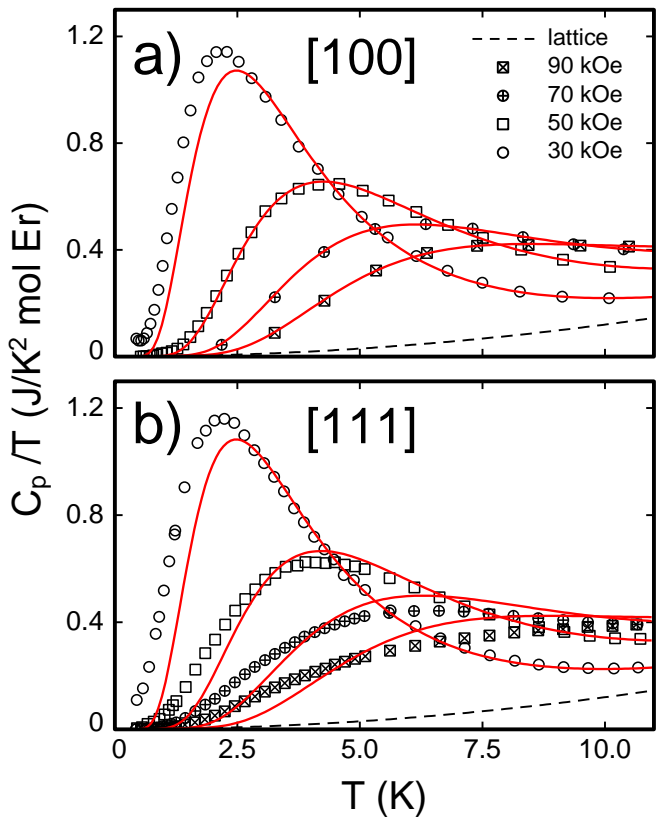}}
\caption{(Colour online). Temperature dependencies of the specific
heat divided by temperature of \ETO\ in various magnetic fields
applied along the $[100]$ (a) and $[111]$ directions (b). Solid
lines represent the fits to equation~(\protect\ref{eqn:schottky})
as described in the main text, the phonon contribution to the
specific heat is shown by the dashed lines.} \label{schott}
\end{figure}

The observed transformation of the $C(T)$ curves on increasing the
external magnetic field up to and above the $H_c$ for these two
field directions is similar to the behavior reported in
Ref.~\onlinecite{gaulin} for $H\parallel [110]$. In higher applied
fields, the sharp low-temperature anomaly seen in the $C(T)$
curves is replaced by a much broader feature developing in the
higher-temperature part of the curves (see Fig.~\ref{schott}).
Following the results of an inelastic neutron scattering
experiment for $H\parallel [110]$ in Ref.~\onlinecite{gaulin}
which showed for the lower excitation mode the dispersion
vanishing at $H\gg H_c$, one can attempt to approximate \ETO\ in
high fields as an ensemble of non-interacting two-levels systems.
In this case, the molar specific heat can be expressed in the
form:
\begin{equation}\label{eqn:schottky}
C_p = \alpha R \left (\frac{\Delta}{kT} \right )^2
\frac{e^{-\Delta /kT}}{(1+e^{-\Delta /kT})^2}+\beta T^3,
\end{equation}
where the first term represents the Schottky anomaly ($\Delta$ is
the gap value, $\alpha\simeq 1$ is a numerical coefficient), and
the second term is the phonon contribution. Taking $\alpha$,
$\beta$ and $\Delta$ as fitting parameters, one can perfectly well
reproduce the specific heat curves in higher fields (50, 70 and
90~kOe) in Fig.~\ref{schott}a for $H\parallel [100]$ over the
entire temperature range shown. For all scans we obtain $\alpha
=0.84\pm 0.01$ and $\beta\simeq (1.33\pm 0.03)\times
10^{-3}$~J/(K$^4$~mole~Er). The latter value agrees well with the
specific heat of an isomorphous nonmagnetic compound $\rm
Y_2Ti_2O_7$ (see {\it e.g.} Ref.~\onlinecite{raju}) rescaled to
\ETO\ by molar mass in the approximation that the Debye
temperature $\theta\propto 1/\sqrt{M}$. The field dependence of
the gap values is presented in the resulting phase diagram
(Fig.~\ref{phasedia}) and discussed below. Only the high
temperature part of the heat capacity measured at $H=30$~kOe can
be satisfactorily described by equation~(\ref{eqn:schottky}), as
the low energy contributions to the entropy are not taken into
account in this simple model. The observed behavior of the
specific heat indicates that the high field excitation spectrum
may indeed be modeled as a set of two-level systems with a
field-dependent gap, while nearer to the critical field, a more
important role is played by the dispersion of the excitation mode.

The specific heat reported by Ruff {\it et al.}~\cite{gaulin} for
a magnetic field along the $[110]$ direction exhibits the same
properties. The high field $C_p(T)$ curves shown in Fig.~1 of
Ref.~\onlinecite{gaulin} can also be described in the manner
outlined above. The corresponding gap value at $H=70$~kOe is shown
on Fig.~\ref{phasedia} for reference.

Using the same model, however, we could not obtain satisfactory
fits to our data obtained for $H\parallel [111]$. The solid lines
on the lower panel of Fig.~\ref{schott} drawn with the same set of
parameters as for $H\parallel [100]$ illustrate the point that
some of the magnetic entropy for this field direction is
redistributed to lower temperatures. One could suggest that the
energy of the long-wave excitations grows more slowly for a
magnetic field applied along the $[111]$ axis than for other field
directions. The relationship between the heat capacity data and
the results of the magnetic resonance measurements is discussed
below.

\begin{figure}
\centerline{\includegraphics[width=\columnwidth]{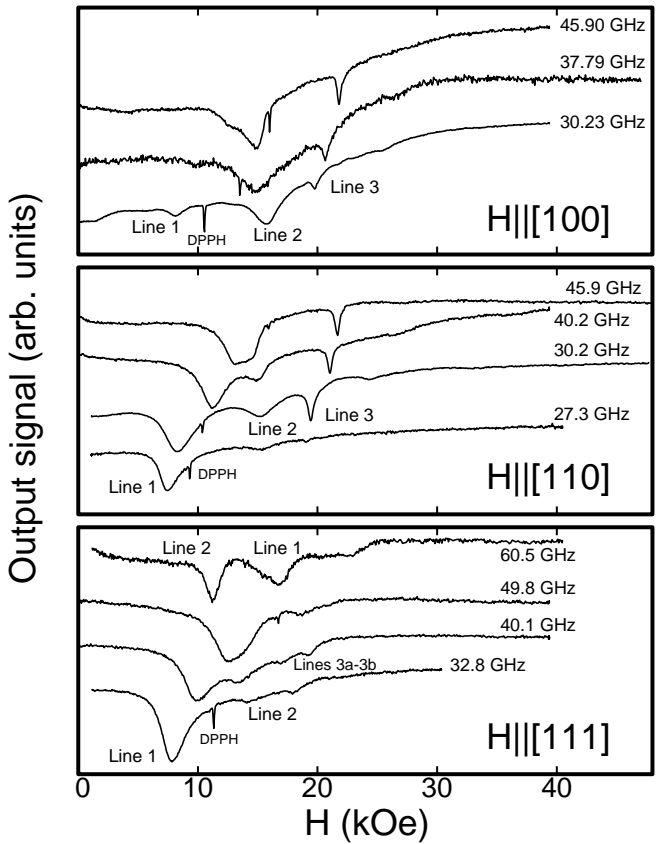}}
\caption{Magnetic resonance absorption spectra in \ETO\ recorded
at the lowest experimental temperature of 0.45~K at different
frequencies of the microwave radiation for the three principal
orientations of the external magnetic field $H\parallel [100]$,
$[110]$ and $[111]$. Narrow weak peaks are the DPPH labels (a
paramagnet with the $g$-factor of 2.} \label{res}
\end{figure}

The resonance absorption spectra recorded at the lowest
experimentally available temperature of 0.45~K at constant
frequency on increasing/decreasing field sweeps are presented in
Fig.~\ref{res}. Rather similar results have been obtained at
$H<H_c$ for fields along the $[100]$, $[110]$ and $[111]$ axes.
Two spectral lines of Lorentzian shape are observed below $T_N$.
The resonance field of line~1 grows linearly with the increasing
frequency while the position of line~2 shifts to lower fields.
Their relative intensities vary with field direction but the
halfwidth of both lines lie in the interval $\Delta H_{1,2}\simeq
1-1.2$~kOe. The nearly frequency independent resonance field of
the second line in the vicinity of the critical field points to
the possible complete softening of this resonance mode, although
our experimental frequency limit does not allow for spectral
records below 25~GHz (corresponding to energies of about 0.1~meV).
At higher fields, for $H>H_c$, a third narrow line ($\Delta
H_3\simeq 0.2$~kOe) appears in the spectra for $H\parallel [100]$
and $[110]$-axes, while for $H\parallel [111]$-axis an absorption
of this type is not seen. Instead, two much broader spectral lines
with considerably smaller intensities (lines 3a,~b on the lowest
panel of Fig.~\ref{res}) are observed. A quantitative analysis of
the frequency-field diagram is performed below.

\begin{figure}
\centerline{\includegraphics[width=\columnwidth]{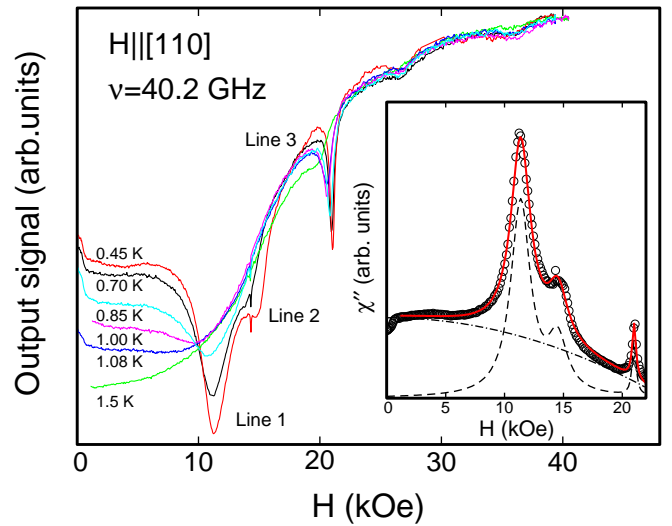}}
\caption{(Colour online) The temperature evolution of the
absorption spectra of \ETO\ measured at the frequency $\nu
=40.2$~GHz for a field parallel to $[110]$. The inset shows the
field dependence of $\chi^{\prime\prime}$ extracted from the
absorption records, fitted by a sum of three Lorentzian spectral
components (dashed line) on the non-resonant background
(dashed-dotted line).} \label{restemp}
\end{figure}

A typical transformation of the resonance spectrum on increasing
the sample temperature above the N\`{e}el point is shown in
Fig.~\ref{restemp}. The records at $\nu=40.2$~GHz reveal an
intensity loss for all three spectral lines on heating. Lines~1
and 2 almost disappear at the transition temperature (about 1.0~K
and 0.8~K for the corresponding resonance fields). Although line~3
exists in the higher fields and is therefore not directly related
to any low-field antiferromagnetic order, it is observed only at
low temperature, broadening and disappearing on warming above
1.5~K. As seen in Fig.~\ref{restemp}, the absorption in the sample
is accompanied by a broad non-resonant background gradually
vanishing in the high field limit. This background signal grows in
amplitude on approaching the N\`{e}el temperature and finally
replaces the three lines of the low-temperature spectrum.
Comparing the spectra recorded at various frequencies one finds
that the background is excited only by the microwave field
component directed along the external magnetic field. Therefore,
the absorption is associated with the longitudinal oscillation of
the magnetic moment and should be damped in the vicinity of the
field at which the magnetization reaches its maximum value. A
similar effect was observed in the Heisenberg pyrochlore system
\GTO\ where the longitudinal susceptibility was supposed to result
from a partial disorder of the magnetic ground state.\cite{sosin0}
As in the present case, this mode had very strong relaxation (of
unknown origin) which makes it impossible to observe other than in
the form of a broad non-resonant background vanishing above the
saturation field. Approximating the system as a set of ``rigid''
sublattices canted to the magnetic field, one obtains the slow
response of the system to the microwave field to be
$\chi\sim\cos^2\varphi$, where $\varphi$ is an angle between the
sublattice and the field. Since the total magnetization of the
system is $M=M_c\sin{\varphi}$ ($M_c$ is a maximum magnetization),
the resulting contribution of such a degree of freedom to the
dynamic susceptibility of the sample can be estimated as
$\chi^{\prime\prime}\propto 1-(M/M_c)^2=1-(H/H_c)^2$ ($H_c$ is a
critical field). The above approach is qualitatively illustrated
in the inset to Fig.~\ref{restemp}. The imaginary part of the
dynamic susceptibility extracted from the absorption spectrum
record was fitted by a sum of three resonance lines of Lorentzian
shape and a non-resonant background with $H_c$ taken as a fitting
parameter. The fit shown by the solid line was achieved for
$H_c=23.6$~kOe which compares reasonably well to the value of the
critical field, given the simplicity of the above approximation.
One should also mention, that although this transition is
associated with slowing down the magnetization growth, it is not a
usual saturation field, as will be discussed below.

\section{Discussion}
For comparison with the experimentally observed spectra one should
first determine the $g$-values of a single Er$^{3+}$ ion in a
crystal field (CF) for the various directions of an external
magnetic field. The total momentum in the ground state of
Er$^{3+}$ ion (electron shell configuration 4f$^{11}$6s$^0$) is
$J=15/2$ ($L=6$, $S=3/2$). The CF of a trigonal $D_{3d}$ point
symmetry at the erbium sites (the three-fold axes are parallel to
local $[111]$ axes) splits the 16-fold state into 8 doublets. A
calculation of this splitting using the CF Hamiltonian parameters
rescaled from another pyrochlore compound Ho$_2$Ti$_2$O$_7$ (given
in Ref.~\onlinecite{rozen}) describes well the observed gaps
between the lowest and two excited CF levels~\cite{champion} and
gives the following expression for the lowest Kramers doublet:
$\textstyle \psi_{1,2}=0.543 \left |\mp\frac{11}{2}\right\rangle
\pm 0.238\left |\mp\frac{5}{2}\right\rangle -0.563 \left
|\pm\frac{1}{2}\right\rangle \mp 0.388 \left
|\pm\frac{7}{2}\right\rangle +0.426 \left
|\pm\frac{13}{2}\right\rangle$. Using these wave-functions, one
can easily calculate the single ion energy splitting under
magnetic field applied at an arbitrary angle $\alpha$ with respect
to a local trigonal axis:
\begin{equation}
\Delta\varepsilon= \mu_BH\sqrt{g_{\parallel}^2\cos^2{\alpha} +
g_{\perp}^2\sin^2{\alpha}}, \label{eq:split}
\end{equation}
where $g_{\parallel}=0.24$ and $g_{\perp}=7.6$ are the $g$-factors
parallel and perpendicular to $[111]$. A more general calculation
which involves the solution of a total atomic Hamiltonian and uses
high temperature susceptibility data in the fitting procedure
gives slightly different lowest doublet wave-functions but the
same $g$-values.\cite{dasgupta} It should be noted that recent
neutron scattering measurements~\cite{Cao} have determined a much
less pronounced XY-anisotropy of the local susceptibility tensor
with the values of $g_{\parallel}=2.6$ and $g_{\perp}=6.8$. When
the field is applied along $[100]$-axis all four magnetic moments
in the unit cell are in equivalent positions with
$\textstyle\cos{\alpha}=1/\sqrt{3}$. Field directions $H\parallel
[110]$ and $[111]$ create two nonequivalent positions with
$\textstyle\cos{\alpha_1}=\sqrt{2/3}$,
$\textstyle\cos{\alpha_2}=0$ and $\textstyle\cos{\alpha_1}=1$,
$\textstyle\cos{\alpha_2}=2\sqrt{2}/3$ respectively. The
corresponding effective $\tilde{g}$-values are collected in a
Table~\ref{table}.

\begin{table}
  \centering
\begin{tabular}{|c|c|c|c|} \hline
  & Ref.~\onlinecite{dasgupta,champion} & Ref.~\onlinecite{Cao} & $\nu_3(H)$ \\ \hline
  $[100]$ & 6.2 & 5.8 & 5.4 \\ \hline
  $[110]$ & 4.4, 7.6 & 4.5, 6.8 & 5.4 \\ \hline
  $[111]$ & 0.2, 7.1 & 2.6, 6.5 & 4.6 \\ \hline
\end{tabular}
\caption{The calculated effective $g$-values of a lowest Kramers
doublet splitting in a magnetic field and the slope of magnetic
resonance branch~3 observed above $H_c$ for three principal field
directions.}\label{table}
\end{table}

\begin{figure}
\centerline{\includegraphics[width=\columnwidth]{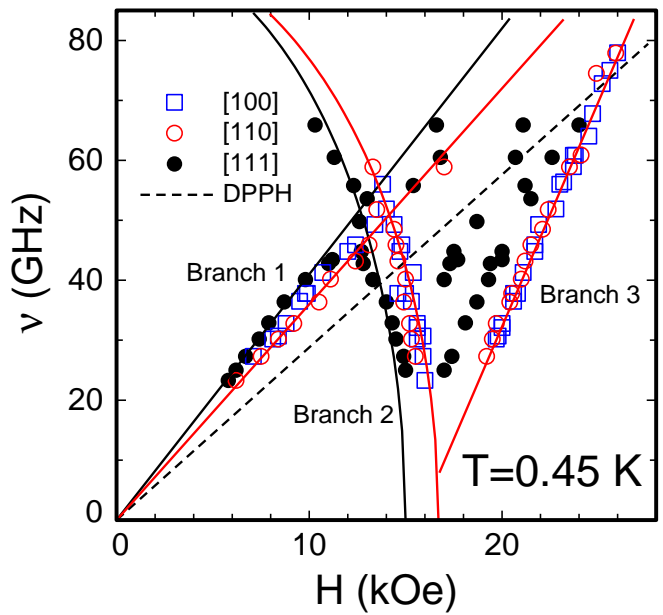}}
\caption{(Colour online) The frequency-field diagram obtained from
the resonance spectra of \ETO\ at $T=0.45$~K (shown in
fig.~\protect\ref{res}): $\square$ -- $H\parallel [100]$, {\Large
$\circ$} -- $H\parallel [110]$, {\Large $\bullet$} -- $H\parallel
[111]$. Fits to all observed spectral modes are shown by solid
lines, dashed line is a DPPH label corresponding to the paramagnet
with $g=2$.} \label{ffd}
\end{figure}

Let us now analyze the resonance spectra measured at the
temperature 0.45~K for the three different orientations of the
external magnetic field. The frequency-field diagram is presented
in Fig.~\ref{ffd}. The resonance spectrum observed below the
critical field $H_c$ consists of two branches. Branch~1 is
increasing in field and can be approximated by a linear dependence
$\nu =\gamma H$ with the value of $\gamma$ slightly dependent on
the field direction: $\gamma_{[100]}=3.8\pm 0.1$~GHz/kOe,
$\gamma_{[110]}=3.6\pm 0.1$~GHz/kOe, $\gamma_{[111]}=4.1\pm
0.1$~GHz/kOe, which correspond to effective $g$-values 2.7, 2.6
and 2.9 respectively. This linear in field gap signifying the
existence of a Goldstone mode at zero magnetic field is quite
unexpected for a strongly anisotropic magnetic system where all
the acoustic modes should be influenced by spin-orbit coupling.
The residual gap due to a possible (within the experimental
accuracy) deviation from the linear extrapolation to zero field
does not exceed a value $\sim 10$~GHz $\simeq 0.04$~meV. One
should also note that the power law dependence of the specific
heat $C_p\propto T^3$ extending to very low
temperatures~\cite{blote,ramirez} confirms the existence of
gapless magnetic excitations at zero field. Obviously, this mode
cannot be interpreted in terms of a single ion splitting. The
observed increase of the gap, which is  linear in field, should be
ascribed to a uniform oscillation of the magnetization resulting
from the in-phase spin motion in the planes perpendicular to the
local $[111]$ axes.

Branch~2 has a gap that diminishes quickly in the vicinity of the
critical field. Although the high frequency measurements were
hindered by the ``size effect'' and the zero field gap of branch~2
is not determined from our experiment, one can relate this branch
to the gapped mode with $\Delta_0\simeq 0.4$~meV observed by
inelastic neutron scattering.\cite{gaulin} The field dependence of
this branch in the vicinity of the second order transition $H=H_c$
is similar to the behavior of the optical branch near the
spin-flip transition: $\Delta\sim\Delta_0\sqrt{1-(H/H_c)^2}$.
Using this formula one can satisfactorily approximate the field
dependence of branch~2 for all three field orientations and
therefore estimate the critical field values: $H_c^{[100]}=16.5\pm
0.5$~kOe; $H_c^{[110]}=16.0\pm 0.5$~kOe and $H_c^{[111]}=15.0\pm
0.5$~kOe. An optical branch with an analogous field dependence was
observed in the spectrum of the Heisenberg pyrochlores \GTO\ and
\GSO. It corresponds to the out-of-phase oscillations of spins in
local ``easy'' planes.

\begin{figure}
\centerline{\includegraphics[width=\columnwidth]{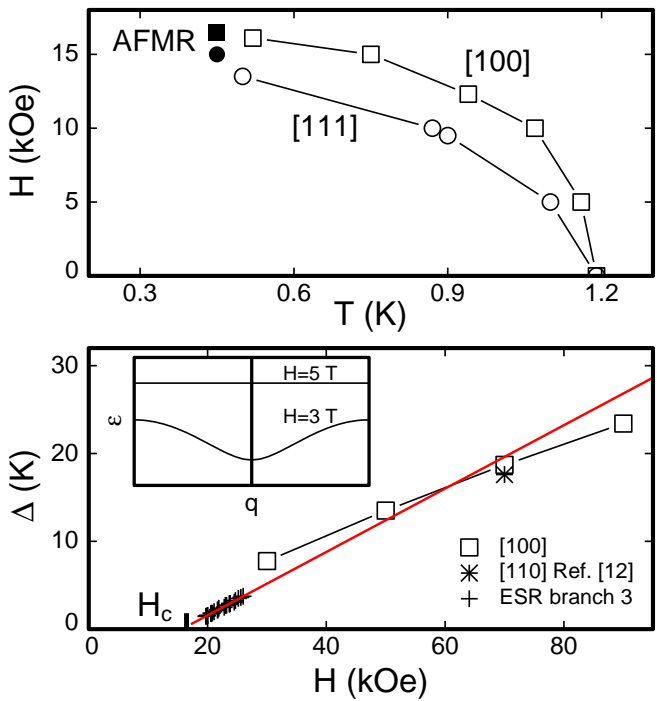}}
\caption{Upper panel: The phase diagram of \ETO\ for $H\parallel
[100]$ ($\square$, $\blacksquare$) and $H\parallel [111]$ ({\Large
$\circ$}, {\Large $\bullet$}) obtained from specific heat and
magnetic resonance data respectively. Lower panel: gap values
determined from fitting the $C_p(T)$ curves for $H\parallel [100]$
by equation~(\protect\ref{eqn:schottky}); $\nu (H)$ dependence of
ESR branch~3 is shown by crosses, solid line is a linear
extrapolation of this branch to high fields. The schematic
transformation of the excitation spectrum at high fields (based on
experimental data from Ref.~\protect\onlinecite{gaulin}) is given
in the inset.} \label{phasedia}
\end{figure}

The analysis of the relative intensities of the magnetic Bragg
peaks~\cite{gaulin} shows that the field-induced transformation of
the magnetic structure can roughly be described as a rotation of
magnetic moments in their local ``easy-planes''. The critical
field is determined by the maximum value of the magnetic moment
which can be achieved without canting from these planes. This
geometric restriction was analyzed theoretically and the
corresponding critical fields were calculated.\cite{glazkov} The
ratio $\textstyle
H_c^{[100]}/H_c^{[110]}=2\sqrt{2}/(\sqrt{3}+1)\simeq 1.04$ is in
agreement with our results within the experimental accuracy while
the theoretical value for $\textstyle
H_c^{[100]}/H_c^{[111]}=7\sqrt{2}/(3\sqrt{3}+\sqrt{6})\simeq 1.3$
appears to be considerably larger than the value 1.1 obtained from
our ESR experiments. In contrast, the corresponding ratio of
critical fields measured by specific heat, $16.5:13.5\simeq 1.2$
demonstrates better agreement with the model. The resulting phase
diagram obtained on the basis of both measurements at $H\parallel
[100]$ and $[111]$ are shown in the upper panel of
Fig.~\ref{phasedia}.

At $H>H_c$, for two directions of the magnetic field $H\parallel
[100]$ and $[110]$, there exists a single gapped resonance mode
that increases linearly in field as $\nu_3 =\Delta (H_c) +
\tilde{\gamma} (H-H_c)$, with roughly the same effective
gyromagnetic ratio $\tilde{\gamma}=7.5$~GHz/kOe ($\tilde{g}=5.4$)
and $\Delta (H_c)\leq 5$~GHz. For $H\parallel [111]$ a resonance
mode of this type is absent and is replaced by two spectral
components with smaller amplitudes and much larger linewidths
giving evidence for increased damping of these oscillations. Their
energies also grow linearly in field with somewhat smaller
$\tilde{\gamma}=6.4$~GHz/kOe ($\tilde{g}=4.6$).

Obviously, the influence of a magnetic field on the collective
excitations above $H_c$ is more associated with the single ion
level splitting than in the antiferromagnetic phase. The linear in
field increase of the gap observed for $H\parallel [100]$ is
rather close to the single ion $g$-factor (especially to the one
reported in Ref.~\onlinecite{Cao}), while for $[110]$ the
$g$-factor lies between the values for two nonequivalent positions
(see Table~\ref{table}). The absence of a well defined collective
oscillation mode in the third field direction $H\parallel [111]$
is most likely the result of a large difference of single ion
$\tilde{g}$-values in two nonequivalent positions which should
destroy the coherent precession of an ensemble of spins with an
average frequency.

One can also establish a relationship between the increase of the
gap as observed directly by the ESR spectroscopy and the evolution
of the specific heat curves under field. The gap values obtained
from the specific heat in the approximation of a single
dispersionless gapped mode (two-level Schottky anomalies) and the
field dependence of branch~3 at $H\parallel [100]$ are presented
in the lower panel of Fig.~\ref{phasedia}. The difference between
them at fields just above the transition results from the effect
of dispersion on the specific heat which is not taken into account
in our fitting. The convergence of the Schottky anomaly gaps to
the high field linear extrapolation of branch~3 is evidence of a
decrease in the dispersion of this mode in the high field limit. A
similar effect, shown schematically in the inset of
Fig.~\ref{phasedia}, was directly observed in inelastic neutron
scattering experiments for $H\parallel [110]$. The agreement
between ESR and specific heat data (single ESR branch and
satisfactory modeling by a two-level system) is also established
for $H\parallel [110]$. In contrast, the failure of such a model
for $H\parallel [111]$ (see Fig.~\ref{schott}) can be considered
as a bulk analog of the absence of a well defined excitation
branch in the ESR measurements.

\section{Summary}
In conclusion, the specific heat and the low temperature magnetic
resonance spectra were studied in the XY pyrochlore
antiferromagnet \ETO. A Goldstone mode with a gap linearly
dependent on the applied field is observed in the ordered state.
Another mode has a zero field gap consistent with the inelastic
neutron scattering results which softens in the vicinity of the
second order transition driven by magnetic field. Presumably,
these collective modes correspond to in-phase and out-of-phase
oscillations of spins in local ``easy'' planes. The critical field
values are determined in three principal field orientations both
from ESR and specific heat measurements. This transition is
associated with the maximum possible spin reorientation without
canting from their planes. In the high field phase the excitation
spectrum for two field directions $H\parallel [100]$ and
$H\parallel [110]$ consists of a single branch. The linear
increase of the gap of this branch in magnetic field is somewhat
related to the $g$-factor values characteristic of a splitting of
a single ion lowest doublet. The specific heat curves observed at
high fields demonstrate a Schottky-type anomaly with a gap that
also increases with field. No well defined excitation mode was
observed above $H_c$ at $H\parallel [111]$. Accordingly, the
specific heat in strong magnetic fields does not demonstrate a
two-level splitting with a single gap. A large difference in
$g$-factors for the two nonequivalent positions of the magnetic
ions with respect to the field is suggested to account for the
absence of a collective excitation mode. Our measurements provide
a framework with which to calculate the excitation spectrum in
\ETO\ using an effective pseudo-spin-$1/2$ model.

\begin{acknowledgments}
The authors thank M.E. Zhitomirsky for fruitful discussions. The
work at Kapitza Institute is supported by the RFBR Grant
10-02-01105. The work at Warwick is supported by the EPSRC Grant
No. EP/E011802. S.S.S is also grateful to LIA LPTMS for the
financial support of the research visit to CEA Grenoble.
\end{acknowledgments}

\end{document}